\documentclass[
	fms,
	times
]{cuparticle}

\usepackage{latexsym,amssymb}

\volume{}
\doi{}

\usepackage[normalem]{ulem}
\usepackage{
	amsmath,
	amsthm,
	amssymb
}
\usepackage{graphicx}
\usepackage{tikz}
\usepackage{tikz-cd}
\usepackage{enumerate}
\usetikzlibrary{matrix}
\usepackage{lscape}
\usetikzlibrary{automata,positioning}
\usepackage{caption}
\usepackage{subcaption}
\usepackage{multicol}
\usepackage{float}
\usepackage{chngpage}
\usetikzlibrary{decorations.pathmorphing}
\tikzset{snake it/.style={decorate, decoration=snake}}
\usepackage{tikz-qtree}
\usepackage{mathtools}
\usepackage{tcolorbox}
\usepackage{collectbox}
\usepackage{pgf}
\usepackage{pgfplots}
\usepackage{longtable}
\usepackage{rotating}
\usepackage{hyperref}

\usepackage{enumerate}

\usepackage{cleveref}

\usepackage[all]{xy}

\newcommand{\la}{\langle}
\newcommand{\ra}{\rangle}
\newcommand{\abs}[1]{\lvert#1\rvert}

\newcommand{\mt}{\mathtt}
\newcommand{\concat}{{}^\frown}

\newcommand{\LPI}{n+1-m-\sum_{i=1}^m (\alpha_i-2) \abs{x_i}\le 2q}
\DeclareMathOperator{\Lookback}{Lookback}

\newtheorem{theorem}{Theorem}
\newtheorem{lemma}[theorem]{Lemma}
\newtheorem{proposition}[theorem]{Proposition}
\theoremstyle{definition}
\newtheorem{definition}[theorem]{Definition}

\begin{document}

\authorheadline{Bj{\o}rn Kjos-Hanssen}
\runningtitle{Incompressibility Theorem for Automatic Complexity}
 
\begin{frontmatter}

\title{An Incompressibility Theorem for Automatic Complexity}
\author[1]{BJ{\O}RN KJOS-HANSSEN}
 
\address[1]{Department of Mathematics, University of Hawai\textquoteleft i at M\=anoa, Honolulu HI 96822
  \ead{bjoern.kjos-hanssen@hawaii.edu}}

\received{25 February 2021}
 
\begin{abstract}
Shallit and Wang showed that the automatic complexity $A(x)$ satisfies $A(x)\ge n/13$ for almost all $x\in{\{\mt{0},\mt{1}\}}^n$.
They also stated that Holger Petersen had informed them that the constant 13 can be reduced to 7.
Here we show that it can be reduced to $2+\epsilon$ for any $\epsilon>0$.
The result also applies to nondeterministic automatic complexity $A_N(x)$.
In that setting the result is tight inasmuch as $A_N(x)\le n/2+1$ for all $x$.
\end{abstract}
 
\end{frontmatter}

	\section{Introduction}

		Kolmogorov's structure function for a word $x$ is intended to provide a statistical explanation for $x$.
		We focus here on a computable version, the automatic structure function $h_x$.
		For definiteness, suppose $x$ is a word over the alphabet $\{\mt{0},\mt{1}\}$.
		By definition $h_x(m)$ is the minimum number of states of a finite automaton that accepts $x$ and accepts at most $2^m$ many words of length $\abs{x}$.
		The \emph{best explanation} for the word $x$ is then an automaton witnessing a value of $h_x$ that is unusually low,
		compared to values of $h_y$ for most other words $y$ of the same length.
		To find such explanations we would like to know the distribution of $h_x$ for random $x$.
		In the present paper we take a step in this direction by studying the case $h_x(0)$, known as the \emph{automatic complexity} of $x$.

		The automatic complexity of Shallit and Wang~\cite{MR1897300} is the minimal number of states of an automaton accepting only a given word among its equal-length peers. Finding such an automaton is analogous to the protein folding problem where one looks for a minimum-energy configuration. The protein folding problem may be NP-complete \cite{protein}, depending on how one formalizes it as a mathematical problem. For automatic complexity, the computational complexity is not known, but a certain generalization to equivalence relations gives an NP-complete decision problem \cite{MR3712310}.

		Here we show (\Cref{thm:july31}) that automatic complexity has a similar incompressibility phenomenon as that of Kolmogorov complexity for Turing machines,
		first studied in~\cite{MR0184801,MR0243922,MR0172744,MR0172745}.

		\subsection{Incompressibility}
		Let $C$ denote Kolmogorov complexity, so that $C(\sigma)$ is the length of the shortest program, for a fixed universal Turing machine, that outputs $\sigma$ on empty input.
		Let $\omega=\{0,1,2,\dots\}$ be the set of nonnegative integers and let $\omega^{<\omega}=\omega^*$ be the set of finite words over $\omega$.

		As Solomonoff and Kolmogorov observed, for each $n$ there is a word $\sigma\in\{\mt{0},\mt{1}\}^n$ with $C(\sigma)\ge n$.
		Indeed, each word with $C(\sigma)<n$ uses up a description of length $<n$, and there are at most $\sum_{k=0}^{n-1} 2^k=2^n-1<2^n=\abs{\{\mt{0},\mt{1}\}^n}$ of those.

		Similarly, we have:
		\begin{lemma}[Solomonoff, Kolmogorov]\label{lem:SK}
			For each nonnegative integer $n$, there are at least $2^n-(2^{n-k}-1)$ binary words $\sigma$ of length $n$ such that $C(\sigma)\ge n-k$.
		\end{lemma}
		\begin{proof}
			For each word with $C(\sigma)< n-k$ we use up at least one of the at most $2^{n-k}-1$ many possible descriptions of length less than $n-k$, leaving
			at least
			\[
				\abs{\{\mt{0},\mt{1}\}^n} - (2^{n-k} - 1)
			\]
			words $\sigma$ that must have $C(\sigma)\ge n-k$.
		\end{proof}

		\subsection{Almost all words of a given length}
		Shallit and Wang connected their automatic complexity $A(x)$ with Kolmogorov complexity in the following \Cref{thm:in_SW_eight}.
		\begin{theorem}[{Shallit and Wang~\cite[proof of Theorem 8]{MR1897300}}]\label{thm:in_SW_eight}
			For all binary words $x$,
			\[
				C(x)\le 12A(x)+3\log_2\abs{x} + O(1). 
			\]
		\end{theorem}
		They mention (\cite[proof of Theorem 8]{MR1897300}), without singling it out as a lemma, the following result (\Cref{lem:almost}).
		Since they used, but did not give a definition of, the notion of \emph{almost all}, we give a definition here. The notion is also known by the phrase \emph{natural density 1}.
		\begin{definition}
			A set of strings $S\subseteq\{\mt{0},\mt{1}\}^*$ contains almost all $x\in\{\mt{0},\mt{1}\}^n$ if
			\[
				\lim_{n\to\infty} \frac{\abs{S\cap \{\mt{0},\mt{1}\}^n}}{2^n}=1.
			\]
		\end{definition}
		\begin{lemma}\label{lem:almost}
			$C(x)\ge \abs{x}-\log_2 \abs{x}$ for almost all $x$.
		\end{lemma}
		\begin{proof} 
			Let $S=\{x\in\{\mt{0},\mt{1}\}^*: C(x)\ge \abs{x}-\log_2\abs{x}\}$.
			By \Cref{lem:SK},
			\[
				\lim_{n\to\infty}\frac{\abs{S\cap \{\mt{0},\mt{1}\}^n}}{2^n} \ge \lim_{n\to\infty}\frac{2^n-(2^{n-\log_2 n} - 1)}{2^n}
				= \lim_{n\to\infty} 1 - \left(\frac1n - \frac1{2^n}\right) = 1.\qedhere
			\]
		\end{proof}
		Shallit and Wang then deduced:
		\begin{theorem}[{\cite[Theorem 8]{MR1897300}}]\label{thm:add_proof}
			For almost all $x\in\{\mt{0},\mt{1}\}^n$ we have
			\(
				A(x)\ge n/13.
			\)
		\end{theorem}
		\begin{proof}
			By \Cref{lem:almost} and \Cref{thm:in_SW_eight} there is a constant $C$ such that for almost all $x$,
			\[
				\abs{x} - \log_2 \abs{x} \le C(x) \le 12A(x)+3\log_2\abs{x} + C.
			\]
			Let $C'=C/12$.
			By taking $n$ large enough,
			\[
				\frac{n}{13}\le \frac{n}{12} -\frac13\log_2 n-C' \le A(x).\qedhere
			\]
		\end{proof}

		Our main result \Cref{thm:july31} implies that for all $\epsilon>0$,
		$A(x)\ge n/(2+\epsilon)$
		for almost all words $x\in\{\mt{0},\mt{1}\}^n$.
		Analogously, one way of expressing the Solomonoff--Kolmogorov result is:
		\begin{proposition}
			For each $\epsilon>0$, the following statement holds: $C(x)\ge \abs{x}(1-\epsilon)$ for almost all $x\in\{\mt{0},\mt{1}\}^n$.
		\end{proposition}

		The core idea for \Cref{thm:july31} is as follows.
		Consider an automaton processing a word $x$ of length $n$ over $n+1$ points in time. We show that there exist powers $x_i^{\alpha_i}$ within $x$ with $\alpha_i\ge 2$, and all distinct base lengths $\abs{x_i}$, that in total occupy $\sum 1+\alpha_i\abs{x_i}$ time, and such that all other states are visited at most twice. Since most words do not contain any long powers, this forces the number of states to be large.

		Automatic complexity, introduced by~\cite{MR1897300}, is an automata-based and length-conditional analogue of
		$CD$ complexity (\cite{Sipser:1983:CTA:800061.808762}).
		$CD$ complexity is in turn a computable analogue of the noncomputable
		Kolmogorov complexity.
		$CD$ stands for ``complexity of distinguishing''. Buhrman and Fortnow \cite{cd-J} call it $CD$, Sipser called it $KD$.
		$KD^t(x)$ is the minimum length of a program for a fixed universal Turing machine that accepts $x$,
		rejects all other strings, and runs in at most $t(\abs{y})$ steps for all strings $y$.

		The nondeterministic case of automatic complexity was studied in~\cite{MR3386523}.
		Among other results, they gave a table of the number of words of length $n$ of nondeterministic automatic complexity $A_N$ equal to a given number $q$ for $n\le 23$, and showed:
		\begin{theorem}[Hyde {\cite[Theorem 8]{MR1897300},\cite{MR3386523}}]\label{thm:Hyde}
			For all $x$, $A_N(x)\le \lfloor n/2\rfloor + 1$.
		\end{theorem}

		In this article we shall use $\la a_1,\dots,a_k\ra$ to denote a $k$-tuple and denote concatenation by $\concat$.
		Thus, for example, $\la 3,6\ra\concat\la 4,4\ra=\la 3,6,4,4\ra$.
		When no confusion is likely we may also denote concatenation by juxtaposition. For example, instead of $U\concat V\concat U\concat C\concat C\concat V$
		we may write simply $UVUCCV$.
		\begin{definition}\label{def:NFA}
			Let $\Sigma$ be finite a set called the \emph{alphabet} and let $Q$ be a finite set whose elements are called \emph{states}.
			A \emph{nondeterministic finite automaton} (NFA) is a 5-tuple
			\(
				M=(Q,\Sigma,\delta,q_0,F).
			\)
			The \emph{transition function} $\delta:Q\times\Sigma\to\mathcal P(Q)$ maps each $(q,b)\in Q\times\Sigma$ to a subset of $Q$.
			Within $Q$ we find the \emph{initial state} $q_0\in Q$ and
			the set of \emph{final states} $F\subseteq Q$.
			As usual, $\delta$ is extended to a function $\delta^*:Q\times\Sigma^*\to\mathcal P(Q)$ by
			\[
				\delta^*(q,\sigma\concat i)=\bigcup_{s\in \delta^*(q,\sigma)}\delta(s,i).
			\]
			Overloading notation we also write $\delta=\delta^*$.
			The set of words accepted by $M$ is
			\[
				L(M)=\{x \in \Sigma^*: \delta(q,x)\cap F\ne\emptyset\}.
			\]
			A \emph{deterministic finite automaton} (DFA) is also a 5-tuple
			\(
				M=(Q,\Sigma,\delta,q_0,F).
			\)
			In this case, $\delta:Q\times\Sigma\to Q$ is a total function and is extended to $\delta^*$ by
			$\delta^*(q,\sigma\concat i)=\delta(\delta^*(q,\sigma),i)$. Finally,
			the set of words accepted by $M$ is
			\[
				L(M)=\{x \in \Sigma^*: \delta(q,x)\in F\}.
			\]
		\end{definition}

		We now formally recall our basic notions.
		\begin{definition}[{\cite{MR3386523,MR1897300}}]
			The \emph{nondeterministic automatic complexity} $A_{N}(x)$ of a word $x\in\Sigma^n$ is
			the minimal number of states of an NFA $M$
			accepting $x$ such that there is only one accepting walk in $M$ of length $n$.

			The \emph{automatic complexity} $A(x)$ of a word $x\in\Sigma^n$ is
			the minimal number of states of a DFA $M$
			accepting $x$ such that $L(M)\cap\Sigma^n=\{x\}$.
		\end{definition}
		Insisting that there be only one accepting walk enforces a kind of unambiguity at a fixed length.
		This appears to reduce the computational complexity of $A_N(x)$, compared to requiring that there be only one accepted word,
		since one can use matrix exponentiation.
		It is not known whether these are equivalent definitions \cite{MR3938583}.

		Clearly, $A_N(x)\le A(x)$. Thus our lower bounds in this paper for $A_N(x)$ apply to $A(x)$ as well.

	\section{The power--complexity connection}\label{sec:aspects}

		The reader may note that in the context of automatic complexity, \Cref{def:NFA} can without loss of generality be simplified as follows:
		\begin{enumerate}
			\item We may assume that the set of final states is a singleton.
			\item We may assume that whenever $q,r\in Q$ and $b_1,b_2\in\Sigma$, if $r\in{\delta(q,b_1)}\cap{\delta(q,b_2)}$ then $b_1=b_2$.
			Indeed, having multiple edges from $q$ to $r$ in an automaton witnessing the automatic complexity of a word would would violate uniqueness.
			\item Each automaton $M$ may be assumed to be \emph{generated by a witnessing walk}. That is, only edges used by a walk taken when processing $x$ along the unique accepting walk need to be included in $M$.
		\end{enumerate}

		Let us call an NFA $M$ \emph{witness-generated} if there is some $x\in\Sigma^*$
		such that $x$ is the only word of length $\abs{x}$ that is accepted by $M$, and $M$ accepts $x$ along only one walk,
		and every state and transition of $M$ is visited during this one walk. In this case we also say that $M$ is witness-generated by $x$.
		When studying nondeterministic automatic complexity we may, without loss of generality, restrict attention to witness-generated NFAs.

		\begin{definition}
			Two occurrences of words $a$ (starting at position $i$) and $b$ (starting at position $j$) in a word $x$ are \emph{disjoint} if
			$x=uavbw$ where $u,v,w$ are words and $\abs{u}=i$, $\abs{uav}=j$.
		\end{definition}

		\begin{definition}
		A \emph{digraph} $D=(V,E)$ consists of a set of \emph{vertices} $V$ and a set of \emph{edges} $E\subseteq V^2$.
		Let $s,t\in V$.
		Let $n\ge 0$, $n\in\mathbb Z$. 
		A \emph{walk} of length $n$ from $s$ to $t$ is a function $\Delta:\{0,1,\dots,n\}\to V$
		such that $\Delta(0)=s$, $\Delta(n)=t$, and $(\Delta(k),\Delta(k+1))\in E$ for each $0\le k<n$.

		A \emph{cycle} of length $n=\abs{\Delta}\ge 1$ in $D$ is a walk from $s$ to $s$, for some $s\in V$, such that
		$\Delta(t_1)=\Delta(t_2), t_1\ne t_2\implies \{t_1,t_2\}=\{0, n\}$.
		Two cycles are \emph{disjoint} if their ranges are disjoint.
		\end{definition}

		\begin{theorem}\label{thm:unlabeled_digraphs}
			Let $n$ be a positive integer.
			Let $D=(V,E)$ be a digraph and let $s,t\in V$.
			Suppose that there is a unique walk $\Delta$ on $D$ from $s$ to $t$ of length $n$,
			and that for each $e\in E$ there is a $t$ with $(\Delta(t),\Delta(t+1))=e$.
			Then there is a set of disjoint cycles $\mathcal C$
			such that
			\[
				v\in V \mathbin{\big\backslash} \bigcup_{C\in\mathcal C} \mathrm{range}(C) \implies \abs{\{t:\Delta(t)=v\}}\le 2,
			\]
			and
			such that for each $C\in\mathcal C$ there exist $\mu_C\ge 2\abs{C}$
			and $t_C$ such that
			\begin{equation}\label{eq:interval}
				\{t: \Delta(t)\in\mathrm{range}(C)\} = [t_C, t_C + \mu_C], \quad\text{and}
			\end{equation}
			\[
				\Delta(t_C+k)=C(k \text{ mod } \abs{C})
				\quad\text{for all $0\le k\le \mu_C$.}
			\]
		\end{theorem}
		\begin{proof}
			Suppose $v\in V$ with $\{t:\Delta(t_j)=v\}=\{t_1<t_2<\dots< t_k\}$ and $k\ge 3$.
			Let us write $\Delta_{[a,b]}$ for the sequence $(\Delta(a),\dots,\Delta(b))$ for any $a,b$.

			\noindent\emph{Claim:} the vertex sequence $S= \Delta_{[t_j,t_{j+1}]}$ does not depend on $j$.

			\noindent\emph{Proof of claim:}
			For $k=3$, $v\in V$ with $\Delta(t_1)=\Delta(t_2)=\Delta(t_3)$, for some $t_1<t_2<t_3$.
			Then the same vertex sequence must have appeared in $[t_1,t_2]$ and $[t_2,t_3]$:
			\[
				\Delta_{[t_1,t_3]}=
				\Delta_{[t_2,t_3]}
				\concat
				\Delta_{[t_1+1,t_2]}
			\]
			or else uniqueness of path would be violated since
			\[
				\Delta_{[0,t_1-1]}\concat \Delta_{[t_2,t_3]}\concat \Delta_{[t_1+1,t_2]}\concat \Delta_{[t_3+1,n]}
			\]
			would be a second walk on $D$ from $s$ to $t$ of length $n$.
			For $k>3$ the only difference in the argument is notational.
			\emph{End of proof of Claim.}

			By definition of the $t_j$'s, $S$ is a cycle except for reindexing.
			Thus, let $C(r)=S(t_1+r)$ for all $r$, let $t_{C}=t_1$, and let
			$\mu=\mu_{C}$ be defined by \eqref{eq:interval}. We have
			\[
				t_C+\mu_C\ge t_k=t_1+\sum_{j=1}^{k-1} t_{j+1}-t_j=t_1 + (k-1)\abs{C}
			\]
			and hence $\mu_C \ge (k-1)\abs{C}\ge 2\abs{C}$.
		\end{proof}

	\section{Main Theorem from power--complexity connection}\label{sec:2}

		\begin{definition}
			Let $\mathbf{w}$ be an infinite word over the alphabet $\Sigma$, and let $x$ be a finite word over $\Sigma$.
			Let $\alpha>0$ be a rational number.
			The word $x$ is said to occur in $\mathbf{w}$ with exponent $\alpha$ if
			there is a subword $y$ of $\mathbf{w}$ with $y = x^a x_0$ where
			$x_0$ is a prefix of $x$,
			$a$ is the integer part of $\alpha$, and
			$\abs{y}=\alpha \abs{x}$. We say that $y$ is an $\alpha$-power.
			The word $\mathbf{w}$ is $\alpha$-power-free if it contains no subwords which are  $\alpha$-powers.
		\end{definition}

		Here in \Cref{sec:2} we show how to establish our Main \Cref{thm:july31}.

		\begin{definition}\label{def:D}
			Let $M$ be an NFA.
			The directed graph $D(M)$ has the set of states $Q$ as its set of vertices and has edges $(s,t)$ whenever $t\in \delta(s,b)$ for some $b\in\Sigma$.
		\end{definition}

		\begin{theorem}\label{cor:simple_implies_square_powers}
			Let $q\ge 1$, $n\ge 0$, and let $x$ be a word of length $n$ such that $A_N(x)= q$.
			Then $x$ contains a set of 
			powers $x_i^{\alpha_i}$, $\alpha_i\ge 2$, $1\le i\le m$, satisfying \eqref{eq:uniqueness} and \eqref{eq:LPI}
			with $\beta_i=\lfloor\alpha_i\rfloor$.
			\begin{equation}\label{eq:uniqueness}
				\sum_{i=1}^m \beta_i \abs{x_i} = \sum_{i=1}^m \gamma_i \abs{x_i},\quad
				\gamma_i\in\mathbb Z, \gamma_i\ge 0
				\implies \gamma_i=\beta_i, \text{for each } i.
			\end{equation}
			\begin{equation}\label{eq:LPI}
			\LPI.
			\end{equation}
		\end{theorem}
		\begin{proof}			Let $M$ be an NFA witnessing that $A_N(x)\le q$.
			Let $D$ be the digraph $D(M)$.
			Let $\mathcal C$ be a set of disjoint cycles in $D$ as guaranteed by \Cref{thm:unlabeled_digraphs}.
			Let $m=\abs{\mathcal C}$ and write $\mathcal C=\{C_1,\dots,C_m\}$.
			Let $x_i$ be the word read by $M$ while traversing $C_i$ and let $\alpha_i=\mu_{C_i}$ from \Cref{thm:unlabeled_digraphs}.

			Since the $C_i$ are disjoint, there are $\Omega := q-\sum_{i=1}^m \abs{x_i}$ vertices not in $\bigcup_i C_i$.
			Let $P=\abs{\{t:\Delta(t)\in C_i, \text{for some }i\}}$ and let $N=n+1-P$.
			By \Cref{thm:unlabeled_digraphs},
			$N\le 2\Omega$ and so $P=n+1-N\ge n+1-2\Omega$.
			On the other hand,
			$P = \sum_{i=1}^m (1+\alpha_i \abs{x_i})$, since a walk of length $k$ is the range of a function with domain of cardinality $k+1$.
			Substituting back into the inequality $P\ge n+1-2\Omega$ now yields
			\[
				\sum_{i=1}^m (1+\alpha_i \abs{x_i}) \ge n+1 - 2\left(q-\sum_{i=1}^m \abs{x_i}\right)
			\]
			and hence \eqref{eq:LPI}.
		\end{proof}

		\begin{theorem}\label{cor:simple_implies_square_powers_unique_lengths}
			Let $q\ge 1$ and let $x$ be a word such that $A_N(x)\le q$.
			Then $x$ contains a set of
			powers $x_i^{\alpha_i}$, $\alpha_i\ge 2$, $1\le i\le m$ such that
				all the $\abs{x_i}, 1\le i\le m$ are distinct and nonzero,
			and satisfying \eqref{eq:LPI}.
		\end{theorem}
		\begin{proof}
			This follows from \Cref{cor:simple_implies_square_powers} once we note that unique solvability of \eqref{eq:uniqueness} implies that
			all the lengths are distinct.

			The unique solution is $\beta_k=\lfloor \alpha_k\rfloor\ge 1$.
			Suppose $\abs{x_i}=\abs{x_j}$, $i\ne j$.
			Then another solution is $\gamma_k = \beta_k$ for $k\not\in\{i,j\}$, $\gamma_i=\beta_i-1, \gamma_j = \beta_j+1$.
		\end{proof}

		For a word $x=x_1\dots x_n$ with each $x_i\in\{\mt 0,\mt 1\}$ we write $x_{[a,b]}=x_ax_{a+1}\dots x_b$.

		\begin{definition}
			Let $x=x_1\dots x_n$ with each $x_i\in\{\mt 0,\mt 1\}$.
			$\Lookback(m,k,t,x)$ is the statement that $x_{m+1+u}=x_{m+1+u-k}$ for each $0\le u<t$, i.e.,
			\[
				\Lookback(m,k,t,x)
				\iff x_{[m+1:m+t]}=x_{[m+1-k:m+t-k]}.
			\]
		\end{definition}
	    We can read $\Lookback(m,k,t,x)$ as ``position $m$ starts a continued run with lookback amount $k$ of length $t$ in $x$''.

		\begin{theorem}\label{thm:july31}
			Let $\mathbb P_n$ denote the uniform probability measure on words $x\in\Gamma^n$, where $\Gamma$ is a finite alphabet of cardinality at least 2.
			For all $\epsilon>0$,
			\[
				\lim_{n\to\infty}\mathbb P_n\left(\left|\frac{A_N(x)}{n/2}-1\right|<\epsilon\right)=1.
			\]
		\end{theorem}
		\begin{proof}
			Let us write $\log=\log_{\abs{\Gamma}}$ in this proof.
			Let $d=3$, although any fixed real number $d>2$ will do for the proof.
			For $1\le m\le n$ and $1\le k\le m$ let $R_{m,k}=\{x\in\Gamma^n: \Lookback(m,k,\lceil d\log n\rceil,x)\}$.
			By the union bound,\footnote{This part is inspired by an argument in~\cite{247929}.}
			\begin{equation}\label{eq:anthony_quas}
				\mathbb P_n\left(\bigcup_{m=1}^n\bigcup_{k=1}^m R_{m,k}\right)\le
				\sum_{m=1}^n \sum_{k=1}^m \abs{\Gamma}^{-d\log n}= n^{-d}\sum_{m=1}^n m =\frac{n(n+1)}2\cdot n^{-d} =: \epsilon_{n,d}.
			\end{equation}
			By \Cref{cor:simple_implies_square_powers_unique_lengths}, if $A_N(x)\le q$ then $x$ contains powers $x_i^{\alpha_i}$ with all $\alpha_i\ge 2$ and all
			$\abs{x_i}$ distinct and nonzero such that \eqref{eq:LPI}, holds:
			\[
			\LPI
			\]
			Applying this with $q=A_N(x)$,
			\begin{equation}\label{eq:feb10_21}
				n+1-m-\sum_{i=1}^m (\alpha_i-2) \abs{x_i}\le 2A_N(x).
			\end{equation}
			Let $S_i=(\alpha_i-1) \abs{x_i}$ and $S=\sum_{i=1}^m S_i$.
			Using $\abs{x_i}\ge 1$ and \eqref{eq:feb10_21}, we have
			\begin{equation}\label{pong}
				n+1-S \le
				n+1-S - m + \sum_{i=1}^m \abs{x_i} \le 2A_N(x).
			\end{equation}
			Using $\alpha_i\ge 2$, and the observation that \emph{if $m$ many distinct positive integers $\abs{x_i}$ are all bounded by $\lceil d\log n\rceil$, then it follows that $m\le \lceil d\log n\rceil$},
			we have
			\begin{equation}\label{pyraminx}
				\{x:\max_{i=1}^m S_i\le \lceil d\log n\rceil\} \subseteq \{x:\max_{i=1}^m\abs{x_i}\le \lceil d\log n\rceil)\} \subseteq \{x: m \le \lfloor d\log n\rfloor\}.
			\end{equation}
			By \eqref{eq:anthony_quas} (since $S_i$ is the length of a continued run in $x$), we have
			\begin{equation}\label{rubiksedge}
				\mathbb P_n(\max_{i=1}^m S_i\le \lceil d\log n\rceil)
				\ge 1-\epsilon_{n,d}.
			\end{equation}

			Using $S \le m \max_{i=1}^m S_i$, \eqref{pyraminx}, and \eqref{rubiksedge},
			\begin{eqnarray*}
				\mathbb P_n(S\le (\lceil d\log n\rceil)^2)
				&\ge& \mathbb P_n(m \max_i S_i\le
				(\lceil d\log n\rceil)^2)\\
				&\ge& 
				\mathbb P_n(\max_{i=1}^m S_i\le \lceil d\log n\rceil)
				\ge 1-\epsilon_{n,d}.
			\end{eqnarray*}
			So by \eqref{pong},
			\begin{equation}\label{eq:main_result}
				\mathbb P_n \left(A_N(x)\ge \frac{n+1}2-\frac12(\lceil d\log n\rceil)^2\right) \ge 1-\epsilon_{n,d}.
			\end{equation}
			Letting $n\to\infty$ completes the proof.
		\end{proof}

	\textbf{Conflicts of interest:} none.
	\paragraph{Financial support.}
	This work was partially supported a grant from the Simons Foundation (\#704836 to Bj\o rn Kjos-Hanssen).

	\bibliographystyle{plain}
	\bibliography{incompressibility}
\enddocument